\documentclass[bm]{epl_ARC}

\usepackage{ifthen}
\usepackage{ifpdf}

\usepackage{latexsym}
\usepackage{amsmath}
\usepackage{amssymb}
\usepackage{bm}

\ifpdf
\usepackage{graphicx}
\usepackage{epstopdf}
\else
\usepackage{graphicx}
\usepackage{epsfig}
\fi

\newcommand{\tbox}[1]{\mbox{\tiny #1}}
 
\newcommand{\amatrix}[1]{\begin{matrix} #1 \end{matrix}}

\newcommand{\be}[1]{\begin{eqnarray}\ifthenelse{#1=-1}{\nonumber}{\ifthenelse{#1=0}{}{\label{e#1}}}}
\newcommand{\ee}{\end{eqnarray}}

\newcommand{\hide}[1]{}

\newcommand{\mpg}[2][\hsize]{\begin{minipage}[b]{#1}{#2}\end{minipage}}
\newcommand{\putgraph}[2][width=\hsize]{\includegraphics[#1]{#2}}

\begin{document}

%%%%%%%%%%%%%%%%%%%%%%%%%%%%%%%%%%%%%%%%%%%%%%%%%%%%%%%%%%%%%%%%%%%%%%%%%%%%
%%%%%%%%%%%%%%%%%%%%%%%%%%%%%%%%%%%%%%%%%%%%%%%%%%%%%%%%%%%%%%%%%%%%%%%%%%%%

\title{The conductance of a multi-mode ballistic ring: \\ beyond Landauer and Kubo}
\shorttitle{}

\author{Swarnali Bandopadhyay, Yoav Etzioni and Doron Cohen}

\institute{
{\small Department of Physics, Ben-Gurion University, Beer-Sheva 84105, Israel}
\vspace*{-0.3cm}
}

\pacs{}{Europhysics Letters {\bf 76}, 739 (2006)\vspace*{-0.7cm}}

%\pacs{03.65.-w}   {Quantum mechanics}
%\pacs{05.45.Mt}   {Quantum chaos}
%\pacs{73.23.-b}   {Mesoscopic systems}

\maketitle

\begin{abstract}
The Landauer conductance of a two terminal 
device equals to the number of open modes 
in the weak scattering limit. 
What is the corresponding result if we close 
the system into a ring? 
Is it still bounded by the number of open modes? 
Or is it unbounded as in 
the semi-classical (Drude) analysis?  
It turns out that the calculation of 
the mesoscopic conductance is similar 
to solving a percolation problem. 
The ``percolation" is in energy space 
rather than in real space. 
The non-universal structures and the sparsity 
of the perturbation matrix cannot be ignored. 
\end{abstract}

%%%%%%%%%%%%%%%%%%%%%%%%%%%%%%%%%%%%%%%%%%%%%%%%%%%%%%%%%%%%%%%%

The theory for the conductance of {\em closed} mesoscopic rings 
has attracted a lot of interest \cite{rings,debye,IS,loc,wilk,kamenev}.  
In a typical experiment \cite{orsay} 
a collection of mesoscopic rings are driven by 
a time dependent magnetic flux $\Phi(t)$ which creates 
an electro-motive-force (EMF) ${-\dot{\Phi}}$ in each
ring. Assuming that Ohm's law applies, the induced current 
is ${I=-G\dot{\Phi} }$ and consequently Joule's law gives 
\be{1}
\mbox{Rate of energy absorption} 
\ \ = \ \ G\,\dot{\Phi}^2 
\ee
where $G$ is called the conductance.
For diffusive rings the Kubo formula 
leads to the Drude formula for $G$.  
A major challenge in past studies 
was to calculate the weak localization 
corrections to the Drude result, 
taking into account the level statistics  
and the type of occupation \cite{kamenev}. 
It should be clear that these corrections 
do not challenge the leading order Kubo-Drude result.

It is just natural to ask what is the conductance 
if the mean free path $\ell$ increases, 
so that we have a ballistic ring as in Fig.~1, 
where the total transmission is ${g_T \sim 1}$. 
To be more precise, we assume that 
the mean free path $\ell \approx L/(1-g_T)$ 
is much larger than the perimeter $L$ of the ring.
In such circumstances ``quantum chaos" considerations 
become important.  
Surprisingly this question has 
not been addressed so far \cite{kbf}, 
and it turns out that the answer requires 
considerations that go well beyond the 
traditional framework.
Following \cite{kbr} we argue that the 
calculation of the energy absorption in Eq.(\ref{e1}) 
is somewhat similar to solving a percolation problem. 
The ``percolation" is in energy space 
rather than in real space. 
This idea was further elaborated in \cite{slr} 
using a resistor network analogy (Fig.~2). 
As in the standard derivation of the Kubo formula 
it is assumed that the leading mechanism for absorption 
is Fermi-golden-rule transitions. These are proportional 
to the squared matrix elements $|\mathcal{I}_{nm}|^2$ 
of the current operator. 
Still, the theory of \cite{kbr} 
does not lead to the Kubo formula. 
This is because the rate 
of absorption depends crucially on the possibility 
to make {\em connected} sequences of transitions,
and it is greatly reduced by the presence of bottlenecks. 
It is implied that both the structure 
of the $|\mathcal{I}_{nm}|^2$ band profile 
and its sparsity play a major role in the calculation of $G$.

%%%%%%%%%%%%%%%%%%%%%%%%%%%%%%%%%%%%%%%%%%%%%%%%%%%%%%%%%%%%%%%%%%%%%%%%%

The outline of this Letter is as follows: 
(a) We define a model example for which the 
analysis is carried out; 
(b) We make a distinction between 
the Landauer, the Drude and the 
actual mesoscopic conductance.
(c) We calculate the matrix 
elements of the current operator; 
(d) We define an ``averaging" procedure 
that allows the calculation of~$G$.  
The result of the calculation is contrasted 
with that of the conventional Kubo approach.

%%%%%%%%%%%%%%%%%%%%%%%%%%%%%%%%%%%%%%%%%%%%%%%%%%%%%%%%%%%%%%%%
%%%%%%%%%%%%%%%%%%%%%%%%%%%%%%%%%%%%%%%%%%%%%%%%%%%%%%%%%%%%%%%%

% \section{The model}

We regard the ballistic ring (Fig.1) as a set    
of $\mathcal{M}$ open modes, 
and a small scattering region 
that is characterized by its total transmission $g_T$. 
To be specific we adopt a 
convenient network model 
where all the bonds (${a=1,2,...,\mathcal{M}}$)   
have similar length ${L_a \sim L}$. 
The scattering is described by 
\be{2}
\bm{S} = \left( \amatrix{
\epsilon\exp\left(i\,2\pi\,\frac{a\, b}{\mathcal{M}}\right) 
& \sqrt{1-\mathcal{M}\epsilon^2} \delta_{a,b} \\ 
\sqrt{1-\mathcal{M}\epsilon^2} \delta_{a,b} 
& -\epsilon\exp\left(-i\,2\pi\,\frac{a\,b}{\mathcal{M}}\right)} 
\right)
\ee
The transitions probability matrix $\bm{g}$ 
is obtained by squaring the absolute values 
of the $\bm{S}$ matrix elements. 
It is composed of a reflection matrix  
$[\bm{g}^R]_{a,b} = \epsilon^2$
and a transmission matrix 
${[\bm{g}^T]_{a,b} = (1-\mathcal{M}\epsilon^2) \delta_{a,b}}$.
The total transmission is 
$g_T = 1-\mathcal{M}\epsilon^2$. 
If the system were open as in Fig.1c.  
then its Landauer conductance would be  
\be{3}
G_{\tbox{Landauer}} 
= \frac{e^2}{2\pi\hbar} 
\sum_{a,b} [\bm{g}^T]_{a,b}
= \frac{e^2}{2\pi\hbar} \mathcal{M}g_T
\ee
If we had a closed ring and we could assume 
that there is no quantum interference within  
the bonds, then we could use 
the multimode conductance 
formula of Ref.\cite{kbf} 
\be{4}
G_{\tbox{Drude}} \ \ = \ \ \frac{e^2}{2\pi\hbar}
\sum_{a,b} \left[ 2\bm{g}^T / (1{-}\bm{g}^T{+}\bm{g}^R) \right]_{a,b}
\ \ = \ \ \frac{e^2}{2\pi\hbar} \mathcal{M}\frac{g_T}{1-g_T}
\ee
The first expression can be derived 
in various ways: Boltzmann picture formalism;  
semiclassical Kubo formalism; or quantum Kubo 
calculation that employs a diagonal approximation.   
In order to get the specific result for 
our network model we had to invert the 
matrix ${(1{-}\bm{g}^T{+}\bm{g}^R)}$.
% 
% which involves some linear-algebra tricks.  
%
We see that in the limit $g_T \rightarrow 1$
the semiclassical $G_{\tbox{Drude}}$ 
is unbounded, while $G_{\tbox{Landauer}}$ is 
bounded by the number of open modes.

%%%%%%%%%%%%%%%%%%%%%%%%%%%%%%%%%%%%%%%%%%%%%%%%%%%%%%%%%%%%%%%%
%%%%%%%%%%%%%%%%%%%%%%%%%%%%%%%%%%%%%%%%%%%%%%%%%%%%%%%%%%%%%%%%

%\section{Objective}

Our objective is to find the conductance 
of the closed ring in circumstances such 
that the motion inside the ring is essentially coherent 
(quantum interference within the bonds is not ignored): 
As in the traditional linear response theory (LRT) 
it is assumed that the level broadening~$\Gamma$ 
is larger compared with the mean level spacing, 
but otherwise very small semi-classically. 
On the other hand, in contrast to LRT, 
we assume ``mesoscopic circumstance", 
meaning that the environmentally-induced relaxation 
is very slow compared with the EMF-induced rate of transitions. 
An extensive discussion of these conditions 
can be found in \cite{kbr}. 
The calculation of~$G$ is done using the formula
\be{5}
G \ = \ \pi\hbar \,\, \varrho_F^2 \times 
\langle\langle |\mathcal{I}_{nm}|^2 \rangle\rangle
\ = \ \frac{e^2}{2\pi\hbar} \times 
2\mathcal{M}^2 \langle \langle |I_{nm}|^2 \rangle \rangle 
\ \equiv \  
\frac{e^2}{2\pi\hbar} \times 
2\mathcal{M}^2  \mathsf{g}
\ee
where $\varrho_F$ is the density of states 
at the Fermi energy, and $\mathcal{I}_{nm}$ 
are the matrix elements of the current 
operator. For our network system 
${\varrho_F = \mathcal{M} L/(\pi\hbar v_F)}$, 
where $v_F$ is the Fermi velocity.
Furthermore it is convenient to 
write ${\mathcal{I}_{nm}=-i(ev_F/L)I_{nm}}$ 
so as to deal with real dimensionless quantities, 
leading to the second expression.
Eq.(\ref{e5}) would be the Kubo 
formula if  ${\langle\langle...\rangle\rangle}$ 
stood for a simple algebraic average. 
But in view of the percolation-like nature  
of the energy absorption process, 
the definition of ${\langle\langle...\rangle\rangle}$ 
involves a more complicated ``averaging" procedure  
that will be discussed and developed later.

%%%%%%%%%%%%%%%%%%%%%%%%%%%%%%%%%%%%%%%%%%%%%%%%%%%%%%%%%%%%%%%%
%\section{eigenstates}

{\bf The eigenstates:}
Our model system, in the absence 
of driving, is time reversal symmetric.
Consequently the unperturbed eigenfunctions  
can be chosen as real
\be{7}
|\psi\rangle 
=\sum_a 
A_a \sin(kx+\varphi_a) 
\,\otimes|a\rangle.
\ee
For a given $g_T$ we can find  
numerically the eigenvalues 
and the eigenstates, thus 
obtaining a table ${(k_n, \varphi_a^{(n)}, A_a^{(n)})}$ 
with ${n=\mbox{level index}}$.  
%
%\be{8}
%(k_n, \varphi_a^{(n)}, A_a^{(n)})
%\ \ \ \ \ \ \ \ n=\mbox{level index}
%\ee
%
In the limit of small $\epsilon$ 
it is not difficult to derive     
the expressions
\be{9}
k_n &\approx& 
\left(
2\pi \times\mbox{\small integer} 
\pm \frac{1}{\sqrt{\mathcal{M}}} \ \epsilon
\right)
\frac{1}{L_a}
\\ 
\varphi_a^{(n)} & \approx &
-\frac{\pi a^2}{\mathcal{M}}
-\frac{1}{2}kL_a
+\left\{ \begin{array}{cc}\pi/4\\3\pi/4\end{array} \right.
\ee
The numerical results over the 
whole range of $g_T$ values are 
presented in Fig.~3.
By normalization we have  
${\sum_a (L_a/2) A_a^2 \approx 1}$.
The degree of ergodicity of a  
wavefunctions is characterized by  
the participation ratio:
\be{11}
\mbox{PR} \equiv
\left[
\sum_a  \left(\frac{L_a}{2} A_a^2 \right)^2
\right]^{-1}
\approx  1 + \frac{1}{3}(1-g_T) \mathcal{M}
\ee
The approximation in the last equality 
is based on the following observations:  
By definition we have ${\mbox{PR} \approx 1}$
for a wavefunction which is localized on one bond, 
while ${\mbox{PR} \sim \mathcal{M}}$
for an ergodic wavefunction.
In the trivial regime $(1-g_T) \ll 1/\mathcal{M}$
the eigenstates are like those of 
a reflectionless ring, with $\mbox{PR} \sim 1$. 
Once $(1-g_T)$ becomes 
larger compared with $1/\mathcal{M}$, 
first order perturbation theory breaks down, 
and the mixing of the levels is described 
by a Wigner Lorentzian. The analysis is completely 
analogous to that of the single mode case in Ref.\cite{kbr}, 
and leads to $\mbox{PR} \propto (1-g_T) \mathcal{M}$.
This is confirmed by the numerical analysis (Fig.~4).
In practice we have found that the proportionality 
constant is roughly~$1/3$.  
Our interest is focused in the {\em non-trivial} 
ballistic regime 
\be{0}
1/\mathcal{M} \ \ \ll \ \  (1-g_T) \ \ \ll \ \ 1 
\ee
where we have strong mixing of 
levels ($\mbox{PR} \gg 1$), but still the 
mean free path $\ell \approx L/(1-g_T)$  
is very large compared with the 
ring's perimeter ($\ell \gg L$).
It is important to realize that in this regime 
we do not have ``quantum chaos" ergodicity.  
Rather we have $\mbox{PR} \ll \mathcal{M}$ 
meaning that the wavefunctions occupy only a small 
fraction  of the classically accessible phase space.

%%%%%%%%%%%%%%%%%%%%%%%%%%%%%%%%%%%%%%%%%%%%%%%%%%%%%%%%%%%%%%%%
%\section{matrix elements}

{\bf The matrix elements:}
The current operator $\mathcal{I}$ is  
the symmetrized version of $e\hat{v}\delta(\hat{x}-x_0)$, 
where $\hat{v}$ and $\hat{x}$ are the velocity 
and the position operators respectively.
The section through which the current is measured 
is arbitrary and we simply take $x_0=+0$.
Given a set of eigenstates, it is 
straightforward to calculate the matrix elements 
of the current operator (Fig.~5), and to 
get insight into their statistical properties
(e.g. Fig.~6). The scaled matrix elements are 
\be{12}
I_{nm} \approx  
\sum_a \frac{L_a}{2} A_a^{(n)} A_a^{(m)} \sin(\varphi_a^{(n)} - \varphi_a^{(m)}) 
\ee
%
% For $n=m$ we have $I_{nm}=0$ as expected from time 
% reversal considerations. 
% From now on we are interested in $n \neq m$. 
%
Needless to say that small PR of wavefunctions 
implies sparsity of $I_{nm}$. It is also worthwhile 
to point out that there are several extreme cases 
that allow simple estimates: 
The case where $n$ and $m$ are 
localized on different bonds leading to ${|I_{nm}|^2=0}$; 
The case where $n$ and $m$ are nearly degenerate 
states localized on the same wire 
leading to ${|I_{nm}|^2=1}$; 
The case where $n$ and $m$ are 
ergodic and uncorrelated
leading to ${|I_{nm}|^2 \approx 1/(2\mathcal{M})}$; 
Irrespective of this, it is clear that 
by normalization the maximal value 
that can be obtained is ${|I_{nm}|^2=1}$.

{\bf Landauer? Drude?} From Eq.(\ref{e5}) and 
the above discussion we deduce that   
\be{13}
&& G\Big|_{\tbox{ergodic}} \ \ \ = \frac{e^2}{2\pi\hbar} \mathcal{M} 
\\ \label{e14}
&& G\Big|_{\tbox{maximal}} \ \ = \frac{e^2}{2\pi\hbar} 2\mathcal{M}^2
\ee
The first expression suggests agreement 
with the Landauer result if we had  
complete ``quantum chaos" ergodicity, 
while Eq.(\ref{e14}) implies a necessary condition 
for a correspondence with the semiclassical 
result Eq.(\ref{e4}):  
\be{15}
\frac{1}{1-g_T} \ll  \mathcal{M}
\ee
This can be rephrased by saying that 
the ballistic time 
${t_{cl} = (1-g_T)^{-1} \times (L/v_F)}$ should 
be much smaller compared with the Heisenberg 
time $t_H=\mathcal{M} \times (L/v_F)$.
In fact it has been argued \cite{kbr}, 
on the basis of a diagonal approximation,  
that semiclassical correspondence is 
indeed realized in the `Kubo calculation".
By ``Kubo calculation" we mean Eq.(\ref{e5}) 
with algebraic average over the near diagonal 
matrix elements of ${|I_{nm}|^2}$. 
The Kubo calculation might have a physical 
validity in the presence of a strong 
relaxation process that suppresses  
the quantum nature of the dynamics.   
See \cite{kbr} for a detailed discussion of this point.

%%%%%%%%%%%%%%%%%%%%%%%%%%%%%%%%%%%%%%%%%%%%%%%%%%%%%%%%%%%%%%%
%\section{The calculation of the conductance}

{\bf The FGR picture:}
The Hamiltonian in the adiabatic basis is  
$\mathcal{H} \mapsto E_n\delta_{nm} + \dot{\Phi} W_{nm}$ 
where $W_{nm} = i\hbar\mathcal{I}_{nm}/(E_n{-}E_m)$, 
and $-\dot{\Phi}$ is the EMF. The FGR transition rate 
between level~$n$ and level~$m$ is proportional to $|W_{nm}|^2$  
multiplied by a broadened $\delta(E_n{-}E_m)$ 
which we call $F()$. The effective broadening 
of the levels reflects either the power spectrum 
or the non-adiabaticity of the driving.  
After trivial scaling the dimensionless 
transition rates are
\be{16}
\mathsf{g}_{nm} \ \ = \ \ \frac{|I_{nm}|^2}{(n-m)^2} \   
\frac{1}{\gamma} F\left(\frac{n-m}{\gamma}\right)
\ee
%
%
% For the purpose of numerical 
% demonstration we assume $F(r)=\exp(-2|r|)$. 
%
The dimensionless broadening parameter~$\gamma$ 
is identical with $\Gamma/\Delta$ 
of Ref.\cite{kbr,pmc} and with $\hbar\omega_0/\Delta$ 
of Ref.\cite{slr}, where $\Delta$ is the mean 
level spacing. There is an implicit approximation 
in Eq.(\ref{e16}), namely $(E_n-E_m)/\Delta \approx (n-m)$,  
that underestimates the exceptionally large couplings
between pairs of almost degenerated levels.
But this is not going to be reflected in the 
energy absorption rate, since the latter 
is indifferent(!) to large sparse values.

{\bf The calculation of the conductance:}
Given the transition rates $\mathsf{g}_{nm}$ 
we want to calculate the rate of energy 
absorption and hence~$G$ as defined by Eq.(\ref{e1}).
It is most convenient to exploit 
the ``resistor network" analogy of Ref.\cite{slr}(Fig.~2). 
Within this framework~$\mathsf{g}$ of Eq.(\ref{e5})
is simply the {\em resistivity} of the network. 
The practical numerical procedure is as follows:
{\bf \ (i)}~Cut an $N$~site segment out of the network.
{\bf \ (ii)}~Define a vector ${\bm{J}_n (n=1..N)}$ 
whose elements are all zero except the first 
and the last that equal ${\bm{J}_1=+J}$ and ${\bm{J}_N=-J}$.
{\bf \ (iii)}~Solve the Kirchhoff equation    
${\bm{J}_n = \sum_{m} \mathsf{g}_{nm} (\bm{V}_n-\bm{V}_m)}$
for the vector $\bm{V}$. 
{\bf \ (iv)}~Find the overall resistance of
the truncated network ${\mathsf{g}_N = J/(\bm{V}_N-\bm{V}_1)}$.
And finally: {\bf \ (v)}~Define the resistivity 
as $\mathsf{g}^{-1}=\mathsf{g}_N^{-1}/N$. 
For a locally homogeneous network it has been argued in \cite{kbr} that  
${\mathsf{g} \approx \langle\langle (1/2)\sum_m (m-n)^2 \mathsf{g}_{nm} \rangle\rangle}$, 
where the sum over~$m$ reflects 
the addition of resistors in parallel, 
and the harmonic average ${\langle\langle...\rangle\rangle}$ 
reflects the addition of resistors in series. 
This expression can be written simply  
as $\mathsf{g} = \langle\langle |I_{nm}|^2 \rangle\rangle$,   
with the implicit understanding that  
the harmonic average is taken over the near diagonal 
elements of the $\gamma$-smoothed $|I_{nm}|^2$ matrix.

{\bf Numerical results:}
The results of the calculation are presented 
in Figs.~7-8. The calculation has been done 
numerically using the resistor network procedure 
that has been explained in the previous paragraph.
We did not to rely on the harmonic average approximation 
because there are prominent structures 
(notably the strongly coupled nearly degenerate levels) 
that make it a-priori questionable. 
However from the numerics it turns out (not displayed) 
that the harmonic average is doing quite well. 
We mention this fact because it gives an insight for 
the numerical results which are displayed in Fig.~7.

Our numerical results suggest that {\em typically} 
${G <  G_{\tbox{Landauer}}}$. 
For an optimal value of $\gamma$, 
such that~$G$ is maximal, it seems that we still 
have  ${G \lesssim  G_{\tbox{Landauer}}}$.  
It is too difficult to figure out the numerical 
prefactor which is involved in the latter inequality (Fig.~8). 
Our conjecture is that this inequality is true {\em in general}   
(disregarding the prefactor which is of order~1).  
We did not find a mathematical argument 
to establish this conjecture, except the 
very simple case of a single mode ballistic 
ring\cite{kbr} where the calculations 
of~$G$ can be done analytically.

%%%%%%%%%%%%%%%%%%%%%%%%%%%%%%%%%%%%%%%%%%%%%%%%%%%%%%%%%%%%%%%
%\section{Conclusions}

{\bf Conclusions:} 
In this Letter we have studied the mesoscopic 
conductance of a ballistic ring 
with mean free path $\ell \gg L$.
The specific calculation has been done for 
a network model, but all its main ingredients 
are completely {\em generic}. 
Ballistic rings with $\ell \gg L$ are not 
typical ``quantum chaos" systems. 
Their eigenfunctions are not ergodic over 
the whole accessible phase space, and therefore 
the perturbation matrix $\mathcal{I}_{nm}$
is highly structured and sparse. Consequently the 
Kubo formula is no longer valid, and one has to 
adopt an appropriate ``resistor network"  procedure 
in order to calculate the true mesoscopic conductance.  
{\em However}, it should be emphasized that if 
there is either a very effective 
relaxation or decoherence process, 
then the theory that we have discussed does not apply. 
In the presence of strong environmental 
influence one can justify, depending 
on the {\em circumstances} \cite{kbf,kbr}, 
either the use of the traditional 
Kubo-Drude result, or the use of the Landauer result.

%%%%%%%%%%%%%%%%%%%%%%%%%%%%%%%%%%%%%%%%%%%%%%%%%%%%%%%%%%%%%%%%
%%%%%%%%%%%%%%%%%%%%%%%%%%%%%%%%%%%%%%%%%%%%%%%%%%%%%%%%%%%%%%%%

{\bf Acknowledgment:}
Much of the motivation for this work came from 
intriguing meetings of DC during 2004-2005 
with Michael Wilkinson, who highlighted 
the open question regarding the 
feasibility to get $G>G_{\tbox{Landauer}}$ 
in the case of a multimode closed ring. 
We also thank Bernhard~Mehlig, Tsampikos~Kottos 
and Holger~Schanz for inspiring discussions. 
The research was supported by the 
Israel Science Foundation (grant No.11/02). 
% and by a grant from the GIF, the German-Israeli Foundation 
% for Scientific Research and Development.
% and by a grant from the DIP

%%%%%%%%%%%%%%%%%%%%%%%%%%%%%%%%%%%%%%%%%%%%%%%%%%%%%%%%%%%%%%%%
%%%%%%%%%%%%%%%%%%%%%%%%%%%%%%%%%%%%%%%%%%%%%%%%%%%%%%%%%%%%%%%%

%%%%%%%%%%%%%%%%%%%%%%%%%%%%%%%%%%%%%%%%%%%%%%%%%%%%%%%%%%%%%%%%
%%%%%%%%%%%%%%%%%%%%%%%%%%%%%%%%%%%%%%%%%%%%%%%%%%%%%%%%%%%%%%%%

% FIGURES:
% blsfig
% knumeric
% PRdistnu
% imageInm; imageInmzoom
% sigmavsgT; sigmavsgTnormal
% conductance

%%%%%%%%%%%%%%%%%%%%%%%%%%%%%%%%%%%%%%%%%%%%%%%%%%%%%%%%%%%%%%%%%%%%%
% illustration of the model

\newpage

\mpg[0.4\hsize]{
\begin{center}
\putgraph[width=\hsize]{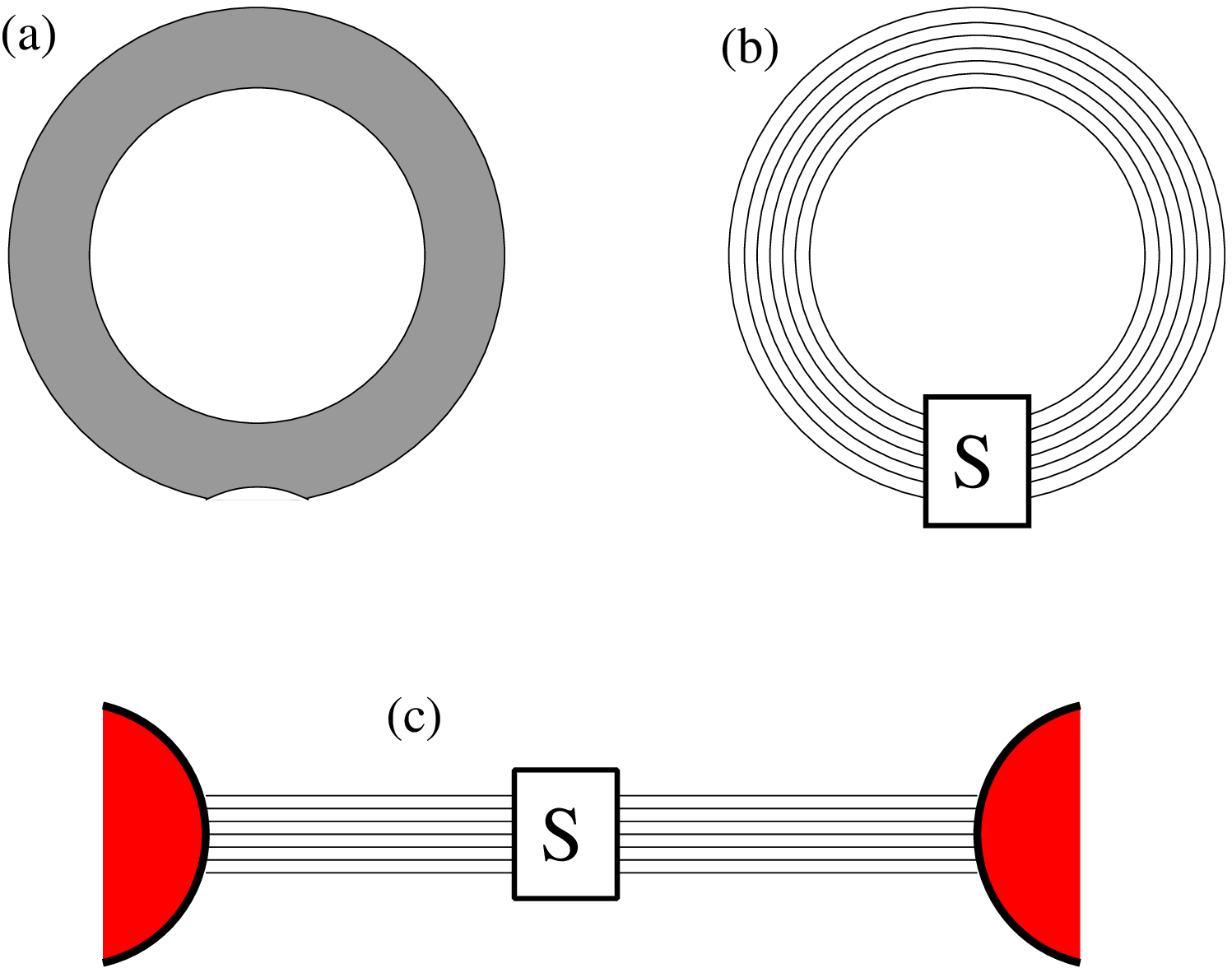}
\end{center}

{\footnotesize 
{\bf Fig.1:} 
{\bf (a)} A billiard example for a ballistic ring. 
The annular region supports $\mathcal{M}$ open modes.  
The electrons are scattered by a small bump. 
{\bf (b)} A network model of a ballistic ring. 
In the numerics the lengths of the $\mathcal{M}$ 
bonds (${0.9<L_a<1.1}$) are chosen in random.
The scattering is described by an $S$ matrix. 
{\bf (c)} The associated open (leads) geometry which 
is used in order to define the $S$ matrix 
and the Landauer conductance. }

}
\hspace{0.1\hsize}
\mpg[0.5\hsize]{
\begin{center}
\putgraph[width=0.9\hsize]{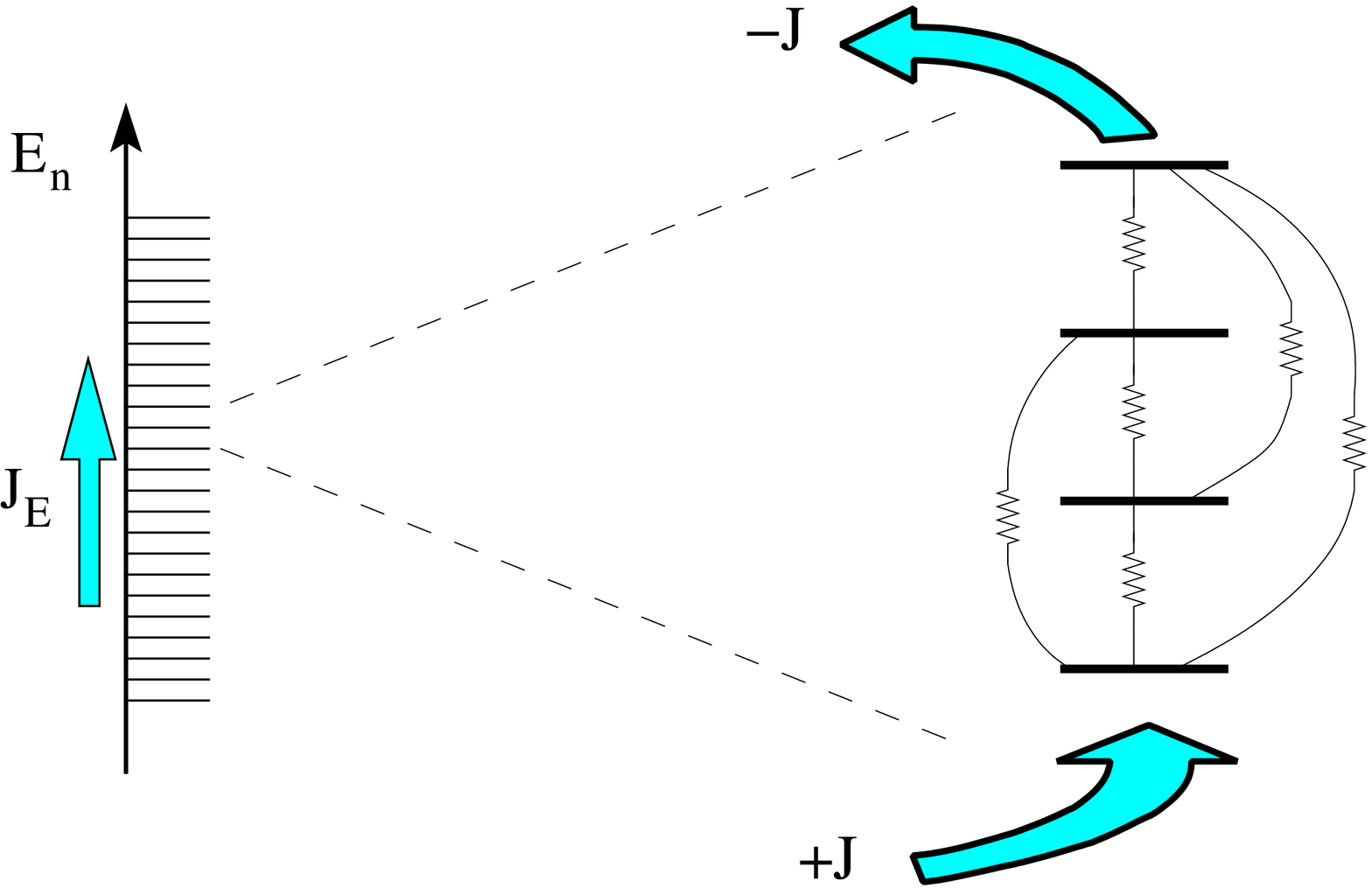}
\end{center}

{\footnotesize 
{\bf Fig.2:} 
The EMF induces diffusion of probability 
in energy space, and hence absorption of energy. 
Within the framework of 
the Fermi golden rule picture the flow 
of the probability in the multi level system 
is analogous to the flow of a current 
via a resistor network. The resistance of 
each ``resistors" corresponds to~$g_{nm}^{-1}$.
The inverse of the diffusion coefficient is   
re-interpreted as the resistivity of the network.
On the right we display a truncated segment, 
where $+J$ is the current injected from one end 
of the network, while $-J$ is the same current 
extracted from the other end. 
The injected current to all the other nodes is zero.} 

}

\mpg[0.45\hsize]{
%%%%%%%%%%%%%%%%%%%%%%%%%%%%%%%%%%%%%%%%%%%%%%%%%%%%%
% eigen-energies versus 1-g

\begin{center}
\includegraphics[clip,width=\hsize]{knumeric}
\end{center}

{\footnotesize 
{\bf Fig.3:} 
The eigenvalues $k_n$ within a 
small energy window around $k \sim 2000$ 
are shown as a function of $1-g_T$.
We consider here a network model 
with $\mathcal{M}=50$ bonds. 
The length of each bond 
was chosen in random  
within $0.9 < L_a <1.1$. } 

}
\hspace*{0.1\hsize}
\mpg[0.45\hsize]{
%%%%%%%%%%%%%%%%%%%%%%%%%%%%%%%%%%%%%%%%%%%%%%%%%%%%%%%
% PRs versus 1-g

\vspace*{8mm}
\begin{center}
\includegraphics[clip,width=\hsize]{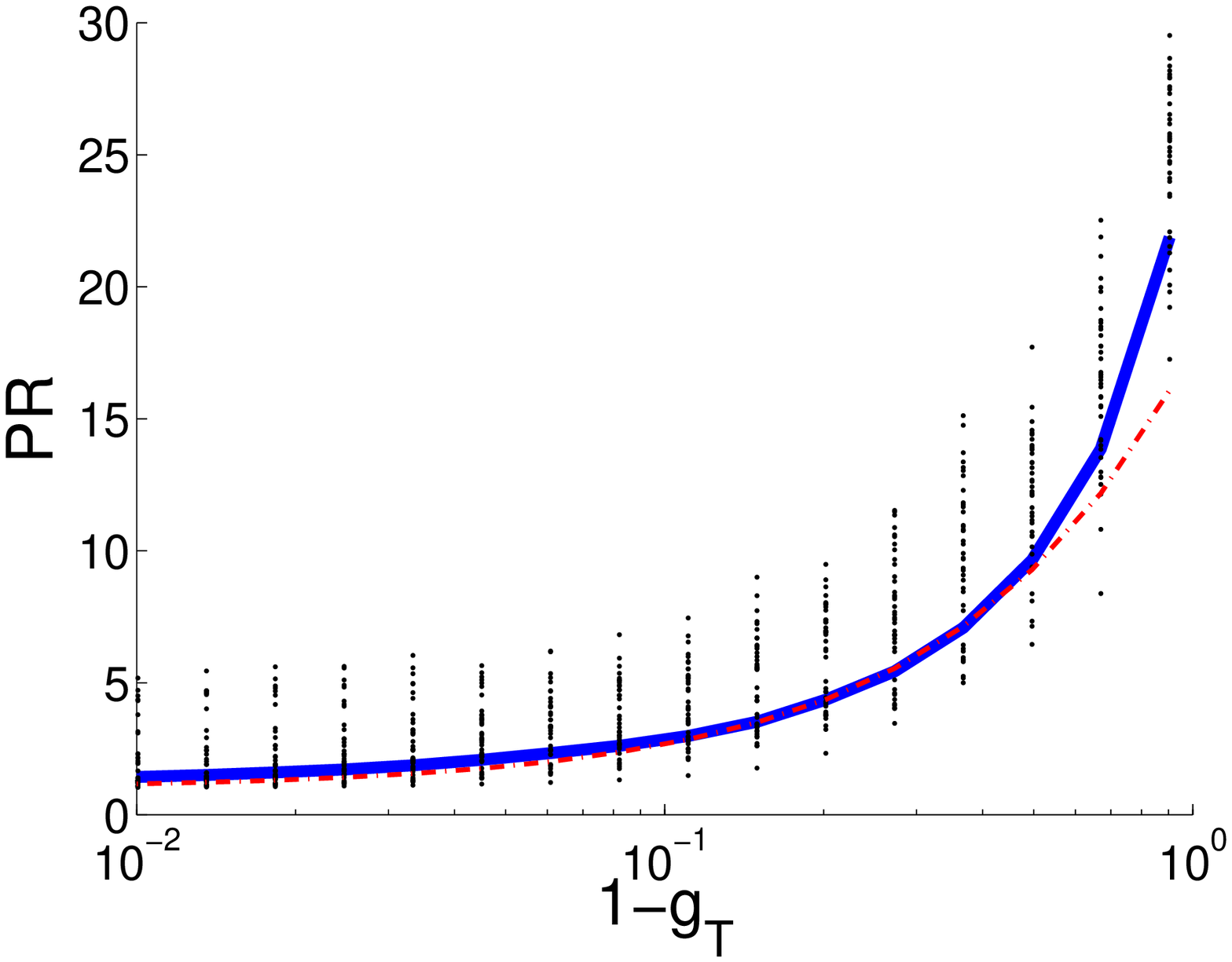}
\end{center}

{\footnotesize 
{\bf Fig.4:} 
For each value of $g_T$ we calculate 
the participation ratio (PR) for all the eigenstates.
We display (as symbols) the minimum value,  
the maximum value, and a set of randomly chosen 
representative values. 
The solid line is the average PR, 
while the dotted line is 
the formula ${\mbox{PR} \approx 1+\frac{1}{3}(1-g_T)\mathcal{M}}$. } 

}

%%%%%%%%%%%%%%%%%%%%%
%%%%%%%%%%%%%%%%%%%%%
\newpage

%%%%%%%%%%%%%%%%%%%%%%%%%%%%%%%%%%%%%%%%%%%%%%%%%%%%%%%%%
% I_nm images

\mpg{
\begin{center}
\includegraphics[clip,width=0.45\hsize]{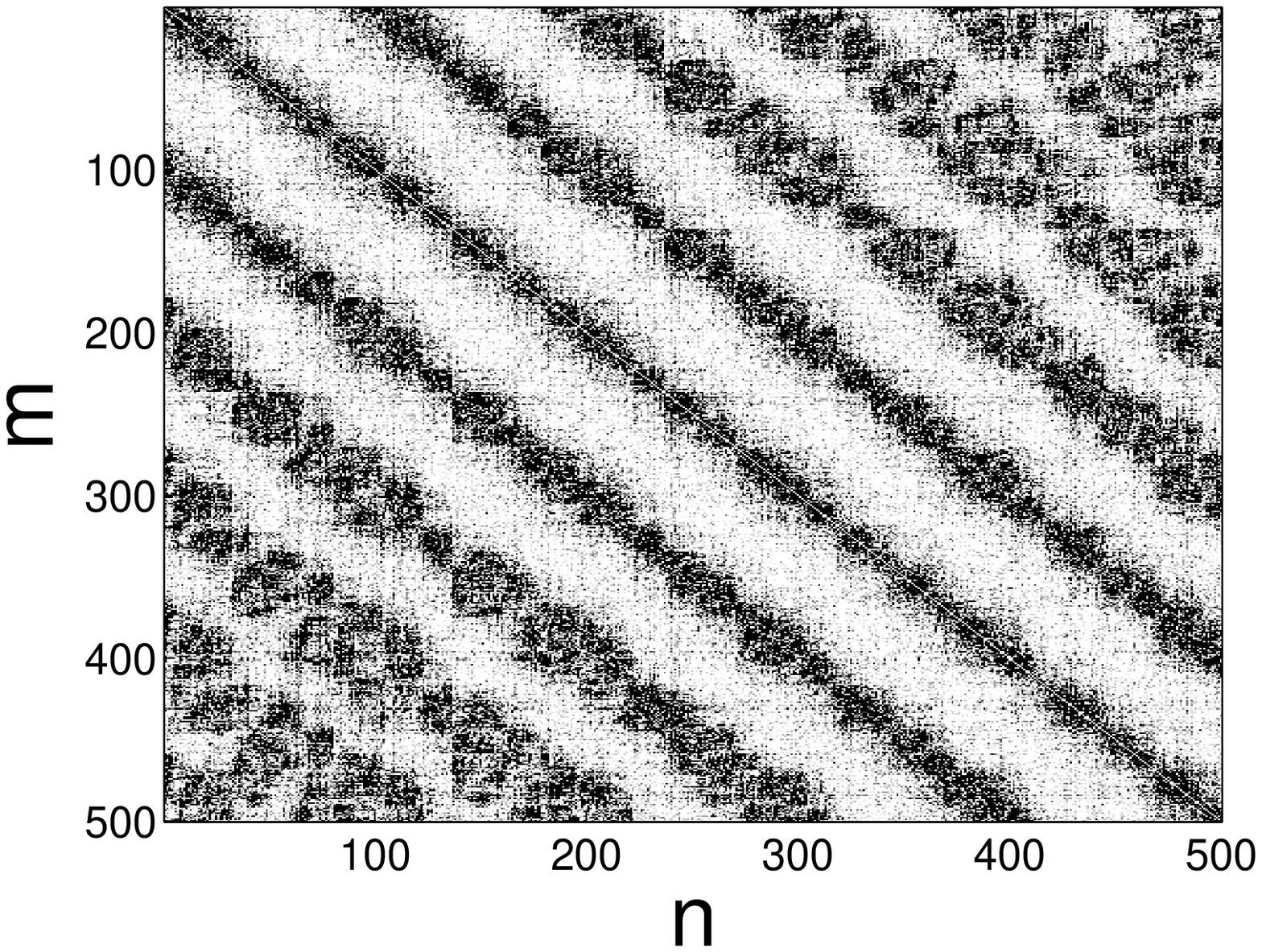}
\includegraphics[clip,width=0.45\hsize]{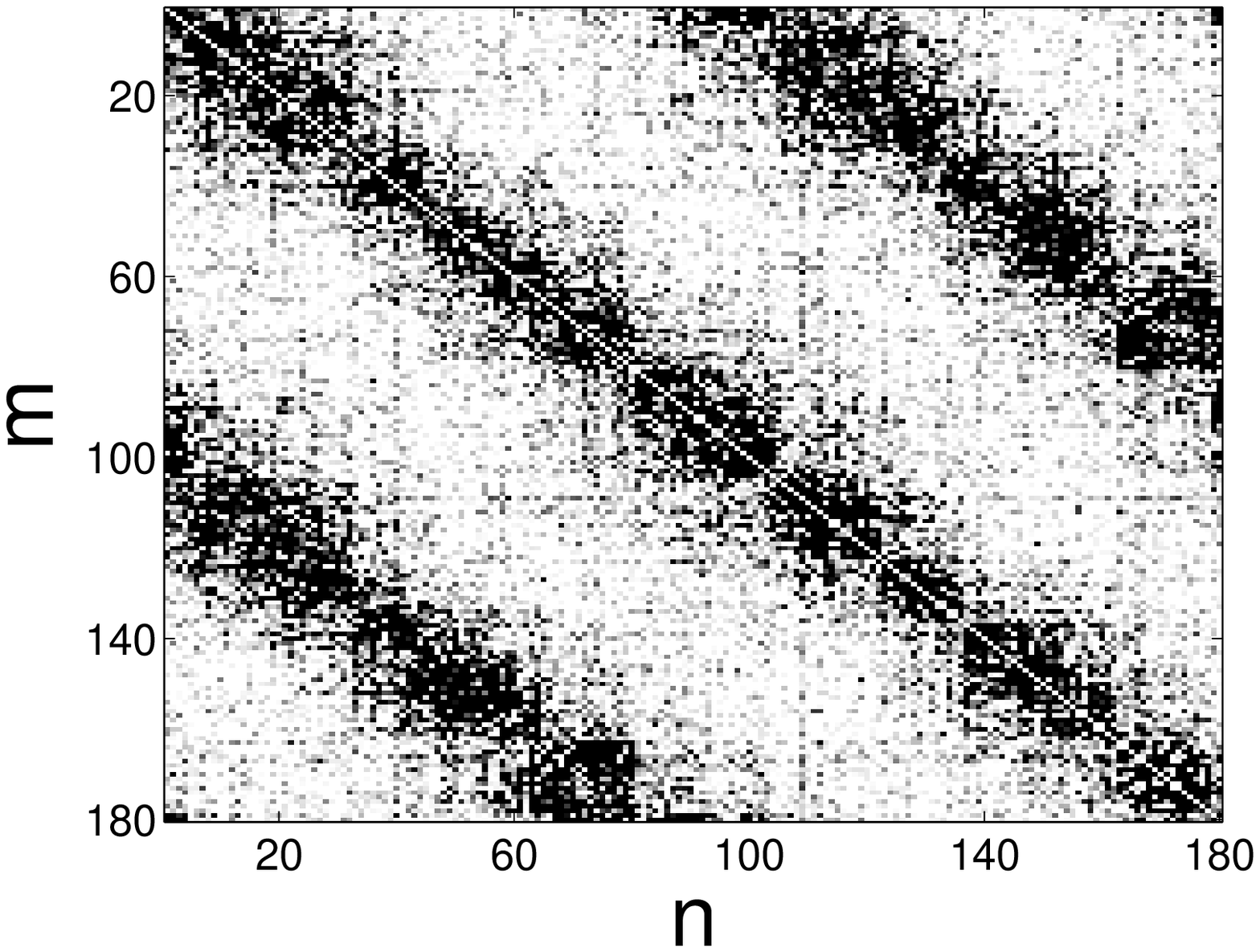}
\end{center}

{\footnotesize 
{\bf Fig.5:}
The image of the perturbation matrix 
$|I_{nm}|^2$ for $g_T=0.9$. 
The right panel is a zoomed image.
If we chose larger $1-g_T$ more elements 
would become non-negligible, 
and the matrix would become less 
structured and less sparse. }

}

\vspace*{5mm}
%%%%%%%%%%%%%%%%%%%%%%%%%%%%%%%%%%%%%%%%%%%%%%%%%%%%%%%%%%
% diagonals

\mpg{
\begin{center}
% corrected figs
\includegraphics[clip,height=0.3\hsize]{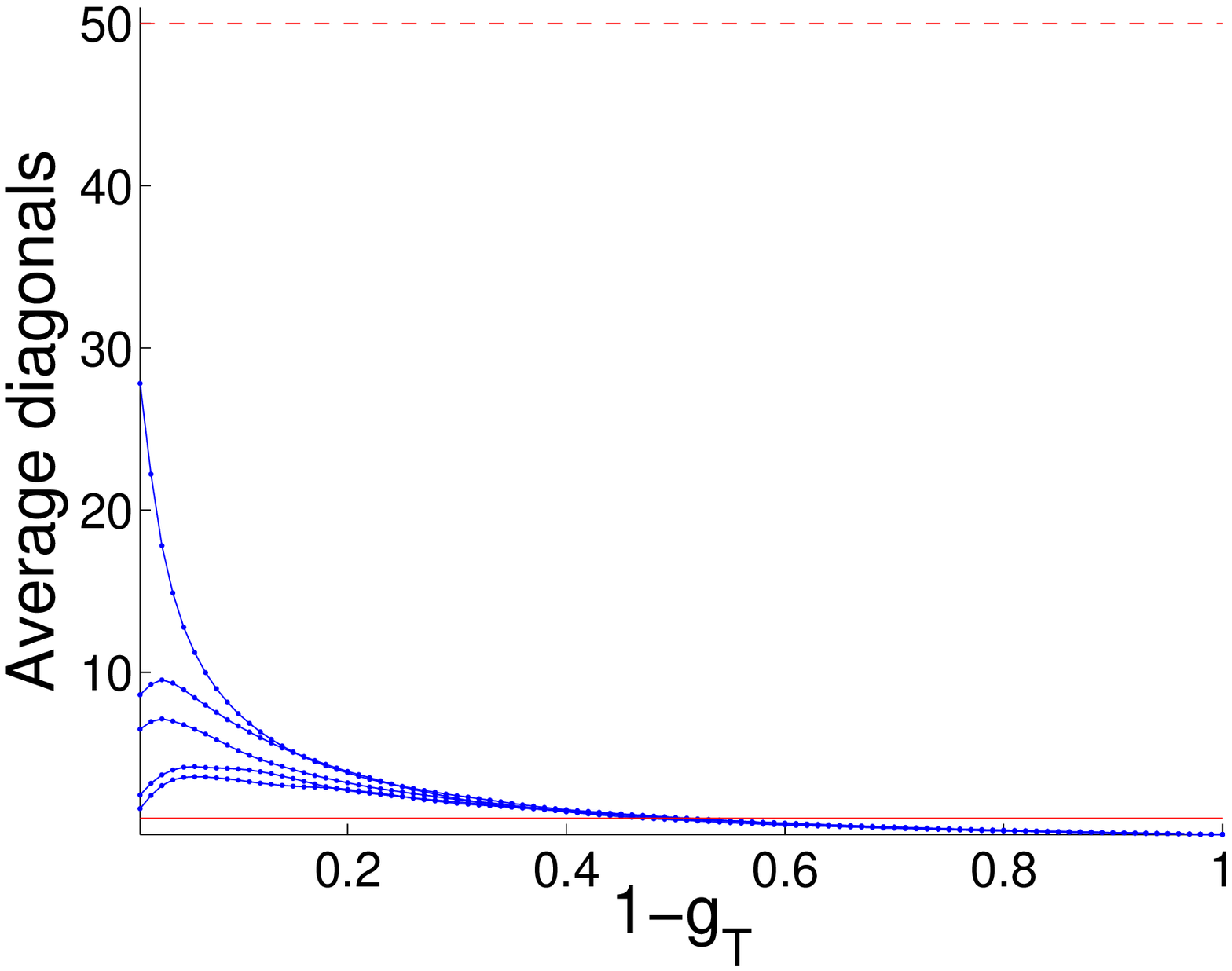}
\includegraphics[clip,height=0.3\hsize]{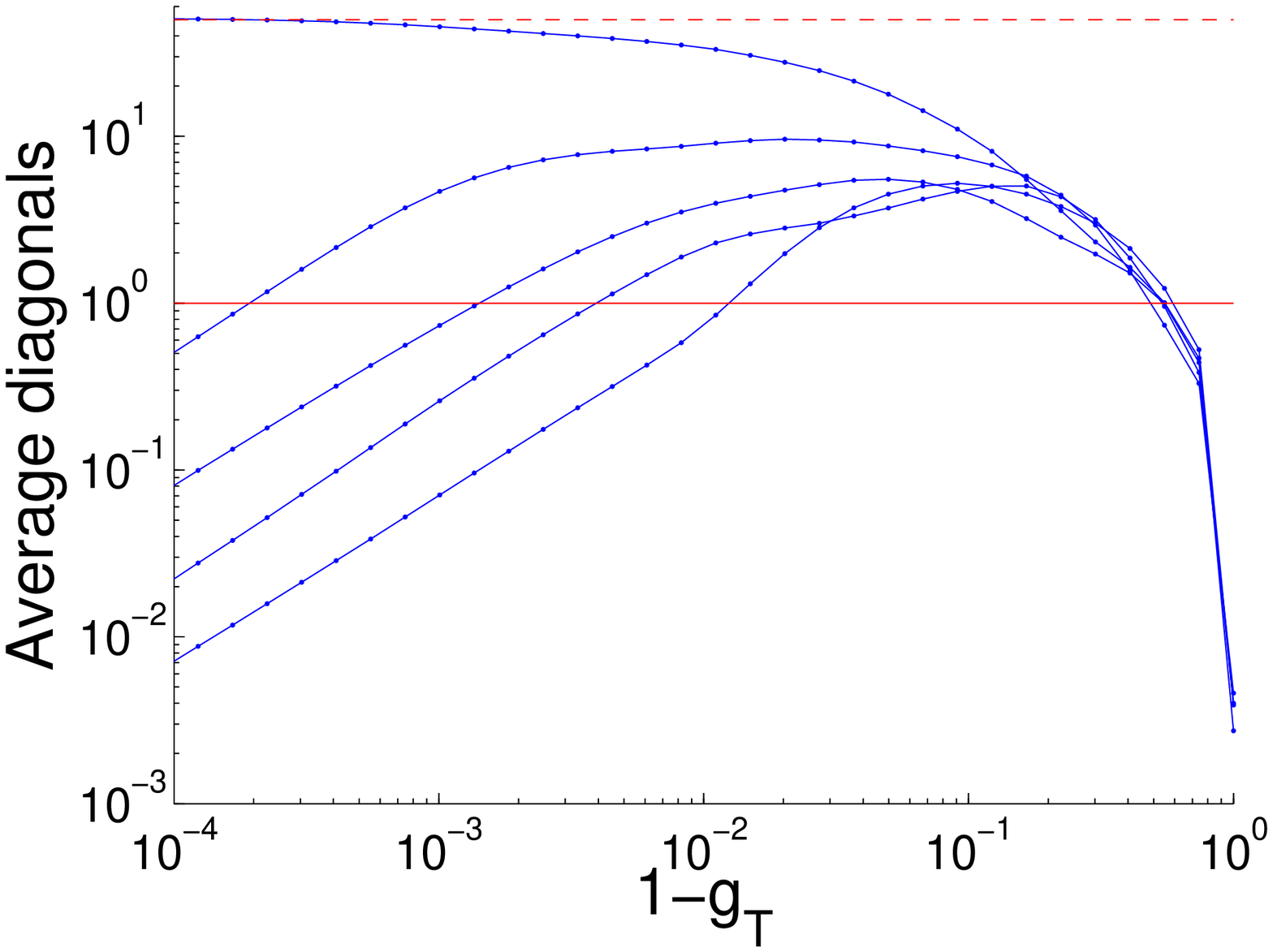}
\end{center}

{\footnotesize 
{\bf Fig.6:}
The $n$-averaged value 
of $2\mathcal{M}^2 |I_{n,n+r}|^2$ 
as a function of $1-g_T$  
for $r=1,2,3,4,5$. 
The ergodic value $\mathcal{M}$ 
and half the maximal value $\mathcal{M}^2$ 
are indicated by horizontal dotted lines.
The left panel is normal scale, 
while the right panel is log-log scale.
}

}

\vspace*{5mm}
%%%%%%%%%%%%%%%%%%%%%%%%%%%%%%%%%%%%%%%%%%%%%%%%%%%%%%%%%%
% conductance

\mpg[0.45\hsize]{

\begin{center}
% corrected fig
\includegraphics[clip,width=0.9\hsize]{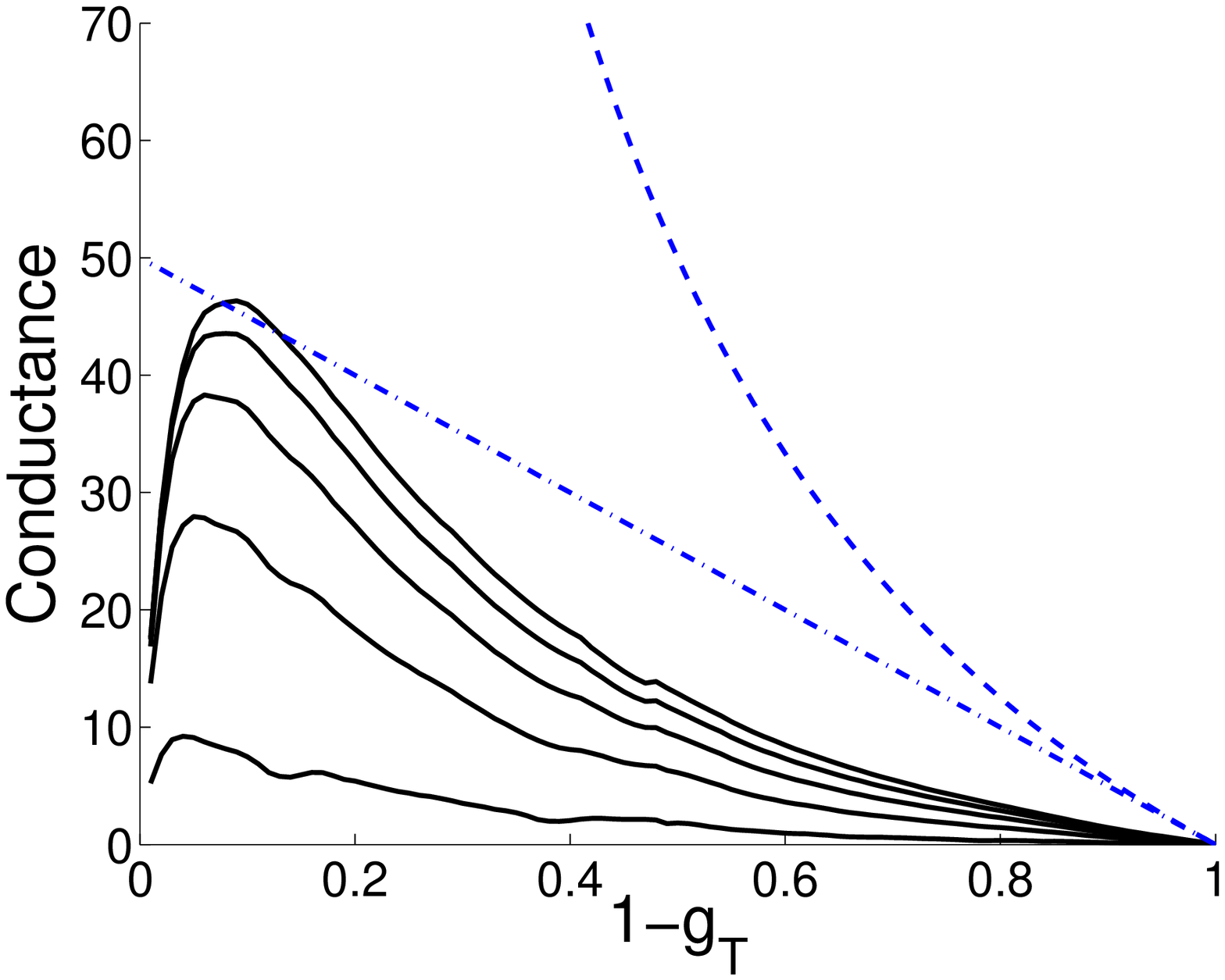}
\end{center}

{\footnotesize
{\bf Fig.7:} 
The mesoscopic conductance $G$ in units of $e^2/(2\pi\hbar)$ 
as a function of $1-g_T$ for $\gamma=1,2,3,4,5$. 
Note again that the total number of open modes 
in our numerics is $\mathcal{M}=50$. The dotted  
line is $G_{\tbox{Landauer}}$ while the dashed  
line is $G_{\tbox{Drude}}$.}

}
\hspace{0.1\hsize}
\mpg[0.45\hsize]{

\begin{center}
\putgraph[width=0.9\hsize]{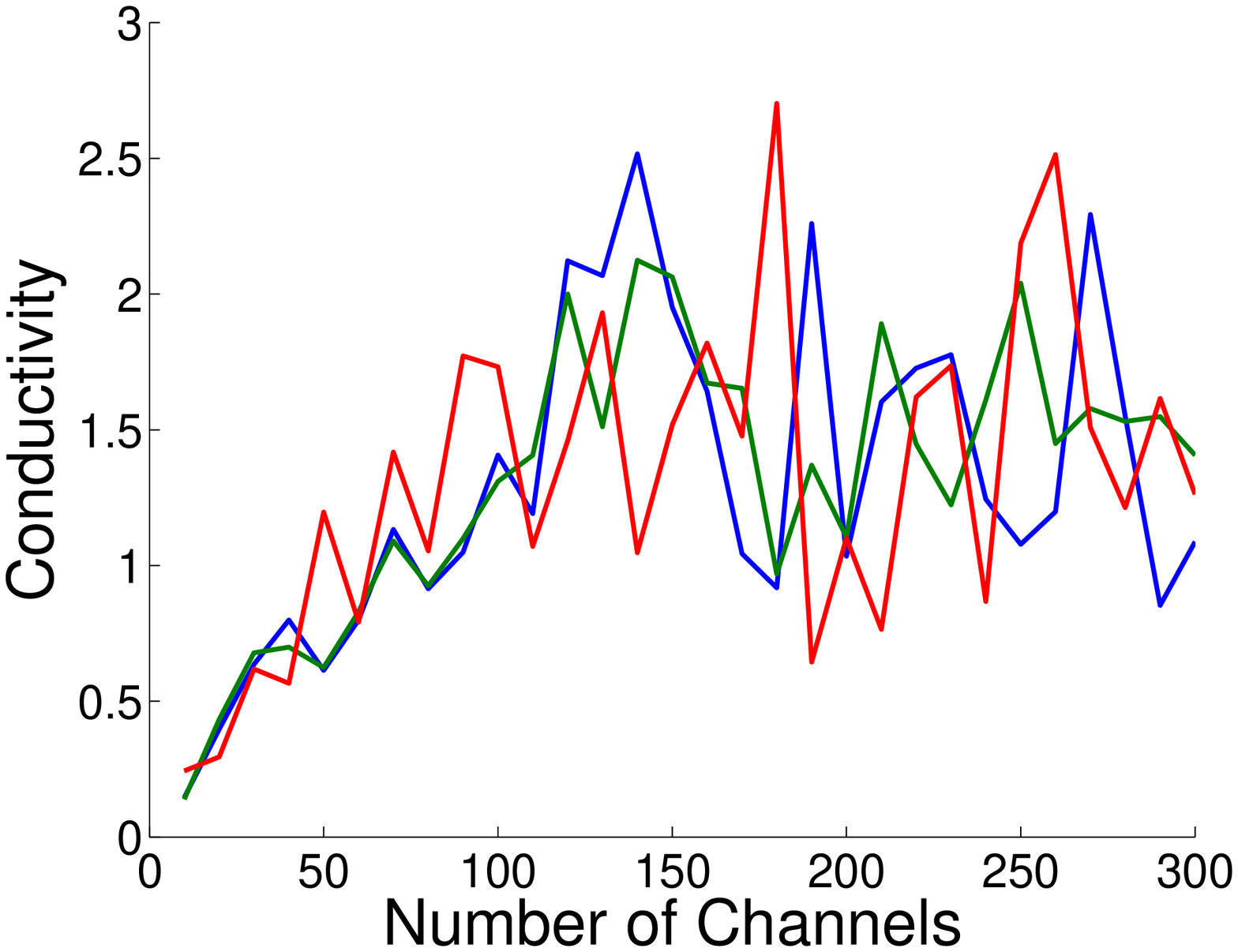}
\end{center}

{\footnotesize {\bf Fig.8:} 
The mesosocopic conductance 
divided by the number of open modes 
for $\mathcal{M}$ up to $300$. 
Here $\gamma=3$ and $g_T=0.8$.
The different curves are calculated 
with segments of length $N=30,50,70$,  
so as to provide an estimate for the numerical error.
}

}

%%%%%%%%%%%%%%%%%%%%%%%%%%%%%%%%%%%%%%%%%%%%%%%%%%%%%%%%%%%%%%%%
%%%%%%%%%%%%%%%%%%%%%%%%%%%%%%%%%%%%%%%%%%%%%%%%%%%%%%%%%%%%%%%%
\end{document}